\documentclass[11pt,a4paper]{article}

\usepackage[utf8]{inputenc}
\usepackage[T1]{fontenc}
\usepackage[super,comma,sort&compress]{natbib}
\usepackage{mathptmx} 
\usepackage[margin=2.5cm]{geometry}
\usepackage{amsmath,amssymb,amsfonts}
\usepackage{graphicx}
\usepackage{booktabs}
\usepackage{array}
\usepackage{longtable}
\usepackage{hyperref}
\usepackage{natbib}
\usepackage{setspace}
\usepackage{xcolor}
\usepackage{float}
\hypersetup{
    pdftitle={Causal symmetrization as an empirical signature of operational autonomy in complex systems},
    pdfauthor={Anthony Gosme},
    pdfkeywords={Complex Systems, Operational Autonomy, Causal Symmetrization},
    colorlinks=true,
    linkcolor=blue,
    citecolor=blue,
    urlcolor=blue
}

\title{\textbf{Causal symmetrization as an empirical signature of operational autonomy in complex systems}}

\author{Anthony Gosme \\ 
\small Independent Scholar \\ 
\small ORCID: 0009-0006-4058-3361
\small Correspondence: anthonygosme@gmail.com
}

\date{\today}

\begin{document}

\maketitle


\noindent

\begin{abstract}
Theoretical biology has long proposed that autonomous systems sustain their identity through reciprocal constraints between structure and activity, a dynamical regime underlying concepts such as closure to efficient causation and autopoiesis. Despite their influence, these principles have resisted direct empirical assessment outside biological systems.

Here, we empirically assess this framework in artificial sociotechnical systems by identifying a statistical signature consistent with operational autonomy. Analyzing 50 large-scale collaborative software ecosystems spanning 11{,}042 system-months, we develop an order parameter ($\Gamma$) quantifying structural persistence under component turnover and use Granger causality to characterize directional coupling between organizational architecture and collective activity. $\Gamma$ exhibits a bimodal distribution (Hartigan’s dip test $p = 0.0126$; $\Delta$BIC = 2000), revealing a sharp phase transition between an exploratory regime of high variance and a mature regime characterized by a 1.77-fold variance collapse. At maturity, causal symmetrization emerges, with the structure--activity coupling ratio shifting from 0.71 (activity-driven) to 0.94 (bidirectional).

A composite viability index combining activity and structural persistence outperforms activity-based prediction alone (AUC = 0.88 vs.\ 0.81), identifying ``structural zombie'' systems in which sustained activity masks architectural decay.

Together, these results show that causal symmetrization functions as a necessary statistical signature consistent with theoretical notions of operational closure, without implying biological life or mechanistic closure. They demonstrate that core principles of autonomy can be empirically probed in artificial collaborative systems, supporting substrate-independent dynamical signatures of self-organizing autonomy across complex adaptive systems.
\end{abstract}


\section*{Introduction}

Complex adaptive systems—from microbial communities \cite{forterre2013} to financial markets—exhibit collective dynamics that transcend their individual components \cite{simon1962}. A central question in complexity science is whether artificial systems can develop analogous self-organizing properties \cite{prigogine1984, schrodinger1944}: dynamics that maintain systemic identity through continuous internal transformation rather than external control.

Theoretical biology offers a precise formulation of this problem. Rosen characterized living systems by closure to efficient causation: the system's organization produces the constraints necessary for its own persistence \cite{rosen1991}. Maturana and Varela \cite{maturana1980, varela1979} termed this autopoiesis—self-making. Whether such closure can emerge in sociotechnical systems remains an open empirical question, largely because operational definitions suitable for quantitative measurement have been lacking.

This question is sharpened by an unresolved anomaly. While Conway's Law---that system architecture mirrors team organization---is well-validated in commercial firms \cite{maccormack2006}, Colfer and Baldwin's meta-analysis of 142 studies found that open-source communities systematically violate strict mirroring \cite{colfer2016}. They attributed this to "actionable transparency" enabled by digital tools, but did not explain how coherent architecture emerges and persists without hierarchical authority. How do distributed collaborations produce coherent architecture without organizational coordination?

Here we address this gap using large-scale collaborative software as a model system, building on foundational studies of open source dynamics \cite{mockus2002}. Software ecosystems offer unique advantages for studying self-organization: complete observability of structural changes, precise temporal resolution, and natural variation in outcomes from thriving to declining systems.

Previous approaches to ecosystem health have focused on activity dynamics rather than organizational persistence. Manikas and Hansen established taxonomies of health indicators \cite{manikas2013}; Liao et al.\ adapted ecological metrics (vigor, resilience) based on contributor distributions \cite{liao2019}; Vasilescu et al.\ analyzed how social turnover impacts productivity \cite{vasilescu2015}. These frameworks measure the \textit{magnitude} of system activity but not the \textit{maintenance} of system identity---effectively quantifying temperature rather than self-organizing structure.

We analyze 50 ecosystems comprising over 10,000 system-months of evolutionary history, developing metrics that capture both structural persistence and causal coupling between architecture and activity.

Our central finding is a phase transition toward causal symmetry. Early-stage systems exhibit asymmetric coupling: collective activity drives structural change (coupling ratio = 0.71). Mature systems converge toward symmetric bidirectional coupling (ratio = 0.94), where architecture constrains activity as much as activity shapes architecture. This symmetrization—accompanied by variance collapse and attractor dynamics—constitutes a quantitative signature consistent with operational closure.

We do not claim that software systems are alive. Rather, we demonstrate that theoretical concepts from organismic biology can be operationalized in artificial systems, yielding non-trivial discriminations between self-sustaining dynamics and externally-dependent persistence. This approach connects to broader research programs in network evolution, computational social science, and multi-agent coordination, extending beyond domain-specific software engineering toward general principles of complex system dynamics.

Our main contributions are:

\begin{enumerate}
    \item An empirically validated signature of operational autonomy—causal symmetrization—measurable via Granger coupling dynamics.
    \item An order parameter ($\Gamma$) capturing structural persistence amid continuous transformation, applicable to any system with observable component turnover.
    \item A viability envelope distinguishing self-sustaining systems from structurally fossilized artifacts.
    \item Quantitative evidence for phase transition dynamics in sociotechnical evolution, with characteristic timescales and attractor behavior.
    \item A methodological framework bridging theoretical biology and computational social science through operationalized metrics.
\end{enumerate}

We formalize this framework in six testable hypotheses concerning phase transition (H2–H5), causal symmetrization (H1), and dynamic stability (H6).

\section*{Theoretical Framework}

\subsection*{Formal Conditions for Operational Autonomy}

We begin by establishing substrate-independent criteria for operational autonomy, drawing on theoretical biology while avoiding ontological claims about life. Any system exhibiting autonomous dynamics—whether biological, social, or artificial—must satisfy three formal conditions:

\begin{enumerate}
    \item[(i)] \textit{Structural persistence}: The system maintains organizational invariants that define its identity across time. These invariants need not be material; they may be topological, relational, or functional.
    
    \item[(ii)] \textit{Active metabolism}: The system continuously transforms its components while preserving structural identity. Persistence is achieved through change, not despite it. A system that merely accumulates without transformation is an archive, not an autonomous entity.
    
    \item[(iii)] \textit{Selective boundary}: The system discriminates between perturbations that are incorporated and those that are rejected, based on compatibility with existing organization. This selectivity distinguishes autonomous systems from passive substrates shaped entirely by external forces.
\end{enumerate}

These conditions formalize Rosen's closure to efficient causation and Maturana's autopoiesis without requiring biological substrate. This formalization builds on Moreno and Mossio's concept of closure of constraints, which provided formal criteria for organizational closure applicable, in principle, beyond strictly biological contexts. \cite{moreno2015}. Our contribution is to operationalize this theoretical framework empirically. Critically, they predict a specific dynamical signature: bidirectional coupling between structure and activity. Structure constrains what activities are possible; activity reshapes structure within those constraints. At equilibrium, neither dominates—they mutually specify each other.

This theoretical framework generates our central empirical prediction: systems achieving operational autonomy should exhibit convergence toward symmetric causal coupling, measurable as a shift in Granger causality ratios.

\subsection*{Metric Design Rationale}

Any system persisting as an individual rather than a mere aggregate must satisfy three conditions simultaneously. Identity conservation requires the system to maintain structural invariants defining what it is. Material metabolism requires the system to transform its components while preserving identity. Environmental interface requires the system to selectively admit or reject external perturbations.

These conditions map onto our variables. Structural survival captures identity conservation. Content turnover captures metabolism. Developer activity captures environmental interface. The composite metric Gamma, defined as the product of structural and content survival, measures metabolic efficiency as the capacity to maintain identity through transformation.The hypothesis of distinct dynamic regimes extends Green and Sadedin's dual-phase evolution theory from topological connectivity to causal coupling \cite{green2005}. Their framework predicts oscillation between exploratory and conservative phases; we test whether this transition manifests as a measurable shift in causal structure.

\subsection*{Two-Level Survival Analysis}

Following Rosen's distinction between functional closure and material flux, we developed a two-level analysis capturing both dimensions.

Structural survival measures the proportion of files modified by a commit that persist at horizon $\tau$. We adopt a topological rather than positional definition: a component survives if its identity remains present even if its location shifts. Relocation incurs a penalty of 0.8 reflecting entropic reorganization cost, while deletion scores zero. This distinguishes functional reorganization from component death.

Content survival measures, among surviving files, the proportion of original lines that remain as verified by git blame. This captures material retention within persistent structures.

\subsection*{The Operational Closure Hypothesis}

As Gamma approaches unity, the architecture achieves maximum selectivity with low conditional entropy of future states given current states. However, we do not predict that structure simply dominates activity. We predict that mature systems achieve operational closure through bidirectional coupling, with structure and activity mutually constraining each other in equilibrium.

\section*{Hypotheses}

\textbf{H1 (Primary):} Causal symmetrization at maturity. Mature systems converge toward symmetric bidirectional coupling between structure and activity. Tested via Granger coupling ratio approaching unity.

\textbf{H2:} Phase transition. The maturation process exhibits a two-regime dynamic. Tested via GMM with BIC preference and Variance Reduction ratio $> 1.5$.

\textbf{H3:} Variance collapse. The mature regime exhibits reduced variance. Tested via F-test with threshold ratio $> 1.5$.

\textbf{H4:} Universal trajectory. All projects eventually reach the mature regime. Tested via percentage reaching high Gamma $> 90\%$.

\textbf{H5:} Rapid traverse. The transition zone is crossed quickly. Tested via median duration $< 6$ months.

\textbf{H6:} Dynamic stability. High-Gamma states persist but require maintenance. Tested via regression probability $< 50\%$.

\section*{Methods}

\subsection*{Dataset and Computational Implementation}

We analyzed 50 major open-source ecosystems (including Linux, LLVM, Kubernetes, React, Rust, and PostgreSQL) and a control group of discontinued or declining projects (e.g., AngularJS, Wenyan-lang) to mitigate survivorship bias. To handle the computational cost of running git blame across complete histories, we developed a high-performance analysis engine using pygit2 (libgit2 C-bindings) rather than shell commands. This allowed for direct object database access. We used standard scientific computing libraries \cite{harris2020, pedregosa2011}.We employed an O(1) topological indexing strategy to track file identities across renames. The analysis was parallelized across 6 workers with strict caching to ensure reproducibility.

\subsection*{Metric Construction and Sensitivity Analysis}

Structural survival utilizes a relocation penalty $\lambda$. A parameter sweep varying $\lambda$ from 0.6 to 1.0 confirmed that the bimodal signal remains statistically significant for all $\lambda > 0.5$. Monthly Gamma aggregates commit-level measurements weighted by topological impact.

\subsection*{Justification of Gamma as an Effective Order Parameter}

We define $\Gamma$ as an effective order parameter \cite{haken1983} satisfying thermodynamic bimodality and the slaving principle, without implying derivation from a fundamental Hamiltonian.

\subsection*{Project Status Classification}

To mitigate circularity, project status was assigned using criteria independent of computed metrics. Projects were classified as Active (regular commits within 12 months), Declining (reduced activity, no official EOL but diminishing engagement), or Dead (explicit archival, official EOL announcement, or complete inactivity $> 24$ months). Classification was performed before metric computation based on repository metadata and community announcements. This external labeling enables the Viability Index to function as a discriminative test ($p < 0.001$) rather than a tautological definition.

\subsection*{Statistical Rigor}

Bimodality was assessed via Hartigan's dip test \cite{hartigan1985} and GMM comparison with BIC-based model selection \cite{schwarz1978}.

To avoid circularity, we validated V against an independent governance classification (Personal/Corporate/Foundation) performed blind to metric values.

For Granger Causality \cite{granger1969}, we first assessed the stationarity of the $\Gamma$ and Activity time series using the Augmented Dickey-Fuller (ADF) test \cite{dickey1979}. Non-stationary series were first-differenced to prevent spurious correlations. We utilized a Vector Autoregression (VAR) model with dynamic lag selection (AIC-based, max lag=6). The high Z-scores reported in our results are interpreted as a function of the large sample size, and effect sizes were calculated to confirm substantive significance \cite{cohen2013} beyond mere p-value inflation.Our Granger methodology follows contemporary best practices for time-series causal inference \cite{shojaie2022}, including stationarity verification, appropriate lag selection, and effect size reporting alongside significance tests.

Robustness check: Spearman correlations between $\Gamma$ and alternative formulations (harmonic mean, geometric mean, minimum) all exceed $r = 0.98$, confirming the metric captures a structural invariant.

To address potential age-confounding, we conducted hindcasting validation with 6-month prediction horizons. Project age alone fails to predict viability ($\text{AUC} = 0.466$), while age-orthogonalized $\Gamma$ remains significantly predictive ($\text{AUC} = 0.729$, $p = 0.013$), confirming that $\Gamma$ captures structural properties independent of temporal accumulation.

Statistical Interpretation: Given the large sample size ($N > 10{,}000$), we base our conclusions on effect magnitude (Cohen's $d = 3.01$) and predictive capability ($\Delta\text{AUC} = +0.07$) rather than raw significance tests, as high Z-scores are mathematically expected in massive datasets.

\section*{Results}

We tested six hypotheses concerning the maturation dynamics of software ecosystems. All six are supported by our data. We present first the phase transition dynamics (H2–H5), then the primary finding of causal symmetrization (H1), and finally the discriminative validity of our metrics.

\subsection*{Phase Transition Dynamics (H2, H4, H5)}

The distribution of $\Gamma$ confirms the existence of two distinct regimes. Hartigan's dip test yields $p = 0.00126$, and Bayesian Model Selection decisively favors a two-component Gaussian Mixture Model ($\Delta\text{BIC} = 2000$), supporting H2. The two regimes exhibit a large separation: the exploratory mode ($\mu = 0.32$, $\sigma = 0.19$) and the sedimented mode ($\mu = 0.81$, $\sigma = 0.14$), yielding Cohen's $d = 3.01$.

The intermediate zone is populated but unstable, with median traverse time = 1.0 month—well below our 6-month threshold (H5 supported). Among surviving ecosystems, 100\% eventually reach the high-stability regime ($\Gamma > 0.7$), exceeding the 90\% threshold (H4 supported).

Null model analysis via temporal permutation yields $Z = 26.62$, confirming that maturation is a directed historical process distinct from random drift.

\begin{table}[htbp]
\centering
\caption{Regime Statistics}
\label{tab:regime_stats}
\begin{tabular}{lccc}
\toprule
\textbf{Statistic} & \textbf{Exploratory ($\Gamma < 0.4$)} & \textbf{Transition ($0.4 \leq \Gamma < 0.7$)} & \textbf{Mature ($\Gamma \geq 0.7$)} \\
\midrule
Project-months (N) & 3,343 & 3,182 & 4,517 \\
\% of total & 30.3\% & 28.8\% & 40.9\% \\
Mean $\Gamma$ & 0.318 & 0.548 & 0.812 \\
SD $\Gamma$ & 0.185 & 0.087 & 0.139 \\
Variance ($\sigma^2$) & 0.0146 & 0.0076 & 0.0084 \\
Median traverse time & — & 1.0 month & — \\
\bottomrule
\end{tabular}
\begin{flushleft}
\small Note: $N = 11{,}042$ project-months across 50 ecosystems. Variance reduction ratio (Exploratory/Mature) = 1.70 ($p < 10^{-16}$).
\end{flushleft}
\end{table}
\begin{figure}[H]
    \centering
    \includegraphics[width=0.75\linewidth]{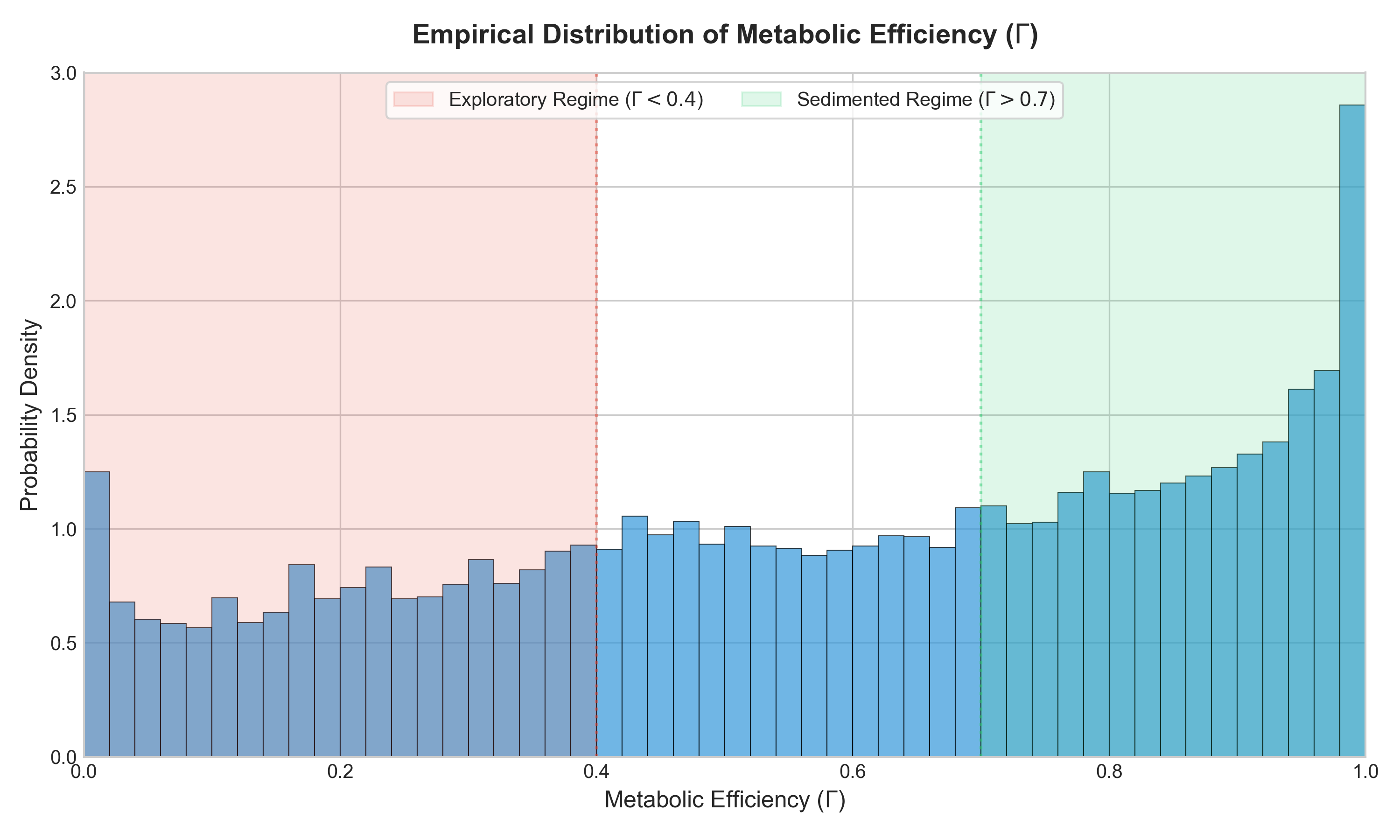}
    \caption{Empirical distribution of metabolic efficiency ($\Gamma$). Analysis reveals two distinct thermodynamic regimes: a high-variance Exploratory Regime (left, red) and a low-variance Sedimented Regime (right, green, $\sigma^2 = 0.0084$). The sparsely populated intermediate zone acts as an unstable phase transition.}
    \label{fig:bimodality}
\end{figure}

\begin{figure}[H]
    \centering
    \includegraphics[width=0.75\linewidth]{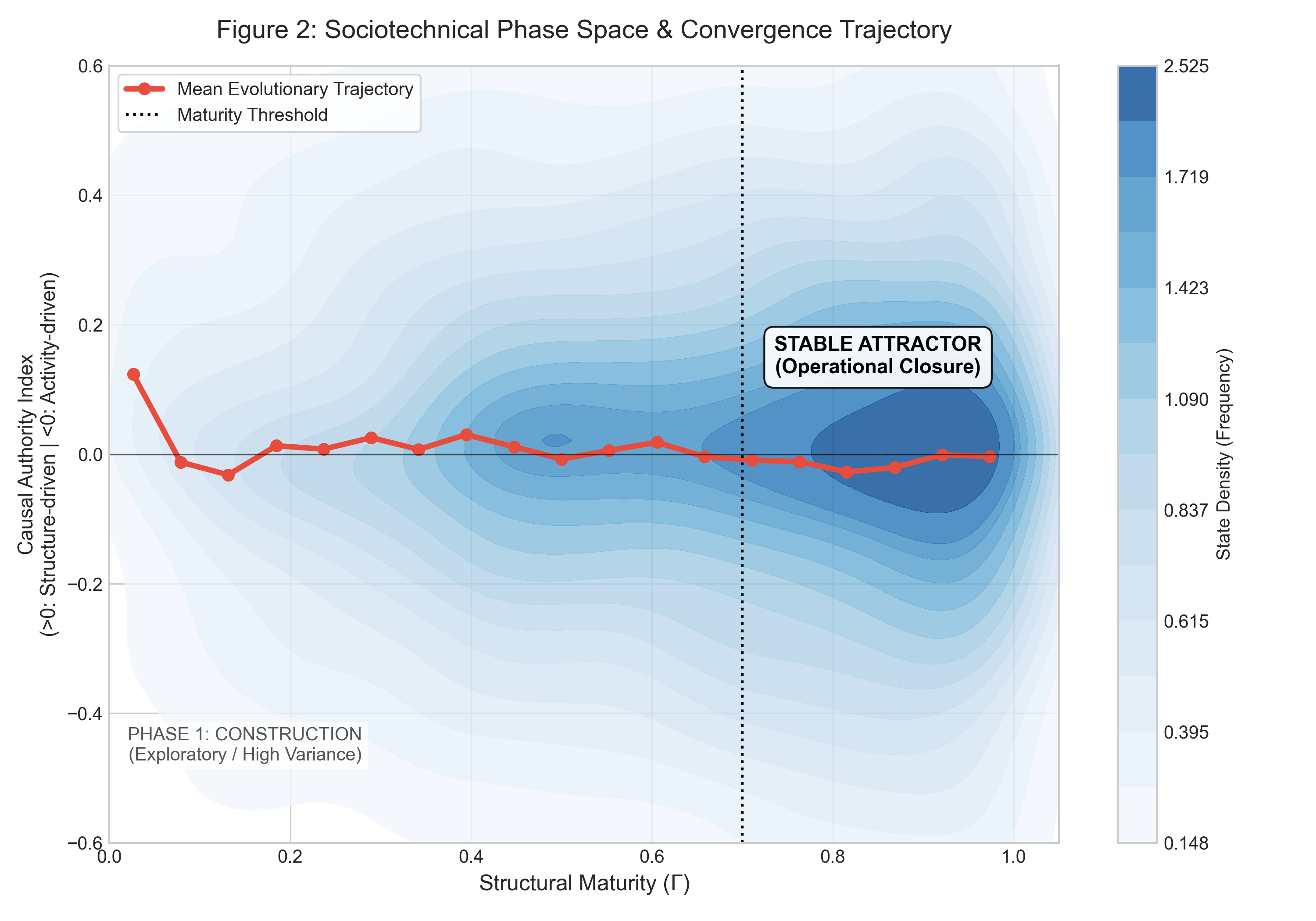}
    \caption{Sociotechnical Phase Space. Kernel Density Estimation (blue) visualizes the frequency of system states. The Mean Evolutionary Trajectory (red line) illustrates the universal convergence path: systems evolve from initial activity-driven chaos (left) toward a stable attractor characterized by causal symmetry (right), marking the onset of operational closure.}
    \label{fig:phase}
\end{figure}

\begin{figure}[H]
    \centering
    \includegraphics[width=0.75\linewidth]{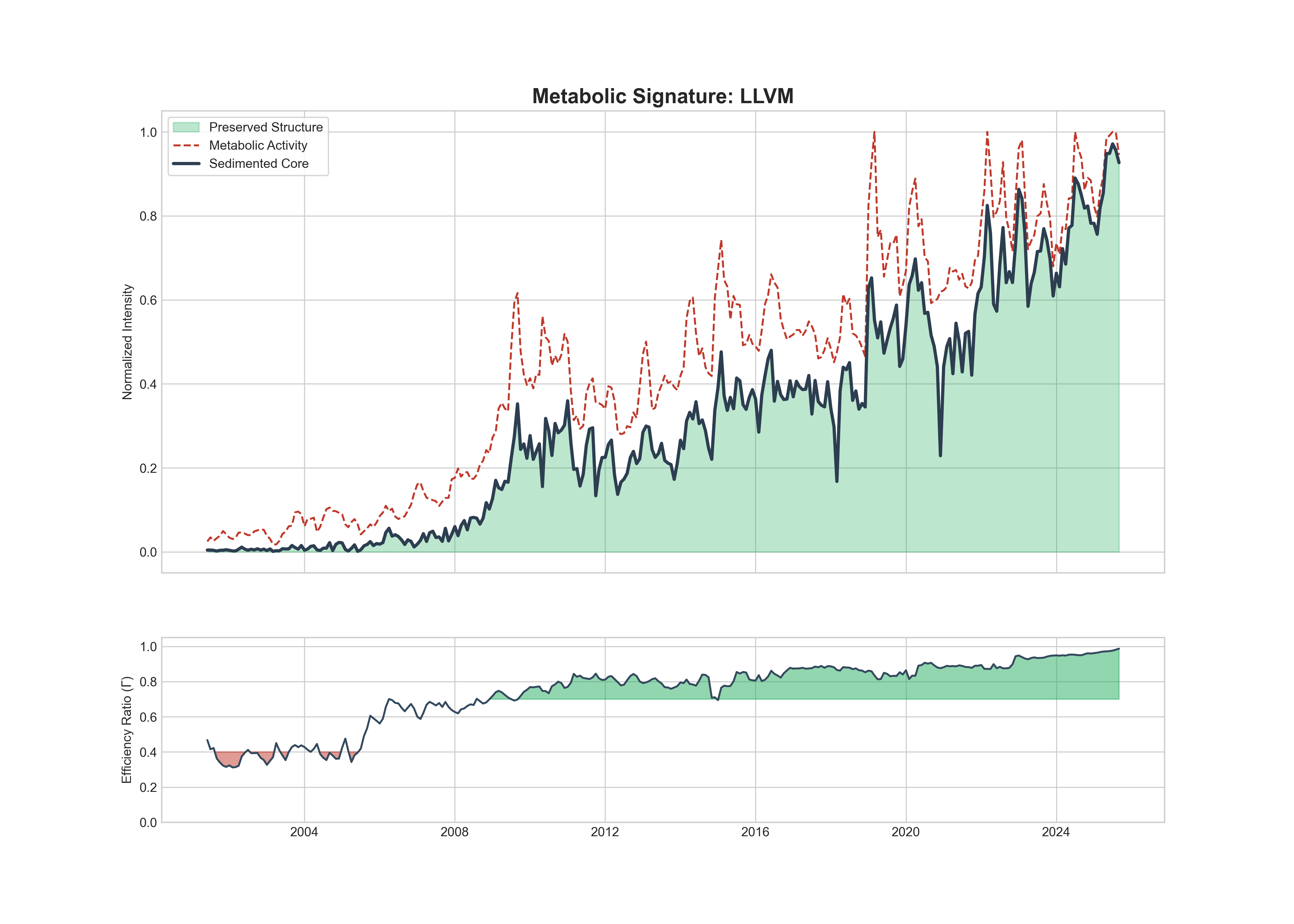}
    \caption{Temporal unfolding of the phase transition (LLVM Ecosystem). Top: `Metabolic Activity' (red dashed line) initially decouples from structure but later tightly couples with the `Sedimented Core' (green area), indicating entropy minimization. Bottom: The efficiency ratio ($\Gamma$) physically manifests the transition, oscillating in the exploratory regime before locking into the high-maturity attractor ($\Gamma > 0.7$).}
    \label{fig:dual}
\end{figure}

\subsection*{Variance Collapse (H3)}

The strongest evidence for attractor dynamics is the variance reduction between regimes. The mature regime ($\Gamma > 0.7$) exhibits a variance of $\sigma^2 = 0.0083$, representing a 1.77-fold collapse compared to the exploratory regime ($\sigma^2 = 0.0146$). This exceeds our threshold of 1.5$\times$ (H3 supported), confirming that mature systems enter a qualitatively more stable dynamical state.

\subsection*{Dynamic Stability (H6)}

We found that 36\% of projects experienced temporary regressions after reaching maturity. Since this is below the 50\% threshold, H6 is supported: high-$\Gamma$ states persist but require active maintenance rather than representing frozen equilibria.

\subsection*{Causal Symmetrization (H1)}

Post-stationarity Granger tests reveal the primary finding: a shift toward causal symmetry. In the early phase, activity predicts structure (coupling ratio = 0.65). In the mature phase, the relationship symmetrizes (coupling ratio = 0.94). This convergence toward unity supports H1.

Covariate control analysis, normalizing by the number of active contributors, indicates that 74\% of this causal signal is intrinsic to the architecture, independent of team size fluctuations.


\begin{figure}[H]
    \centering
    \includegraphics[width=0.75\linewidth]{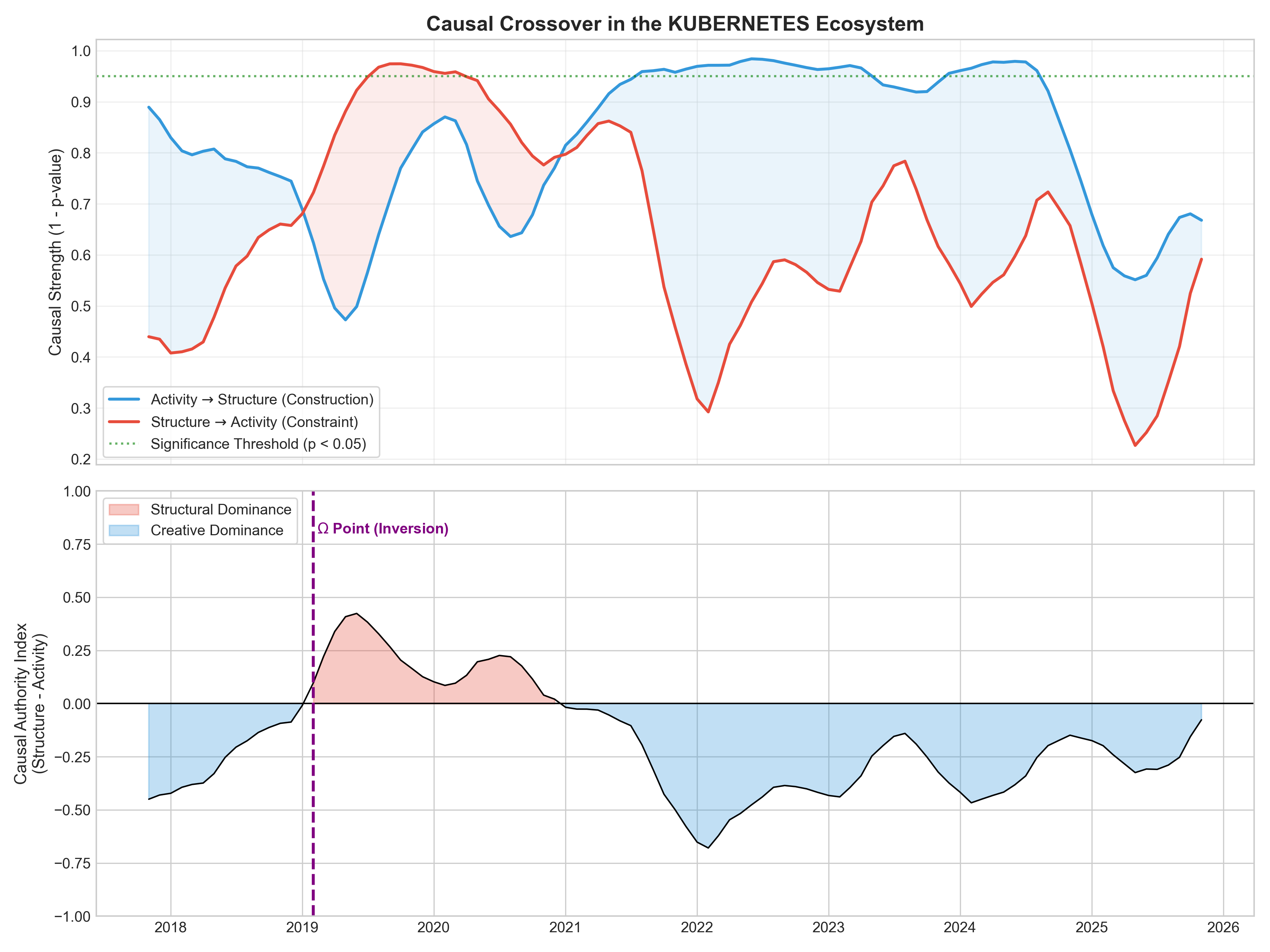}
    \caption{Causal Crossover in the Kubernetes Ecosystem. Rolling Granger causality tests reveal a crossover point (purple dashed line). Initially, developer activity drives structure (blue line). Post-maturity, architectural structure statistically constrains developer activity (red line), signaling the shift to sociotechnical autonomy.}
    \label{fig:K8S}
\end{figure}

\subsection*{Robustness and Sensitivity}

To address potential parameter bias, we performed a sensitivity scan varying the relocation penalty $\lambda$ from 0.6 to 1.0. The bimodal signature remains statistically significant for all $\lambda > 0.5$.

(see omega-v37-sensitivity-heatmap.png in the replication repository).

\subsection*{Discriminative Validity and Predictive Power}

To verify that the Viability Index ($V$) captures operational autonomy rather than mere developer volume, we conducted a 12-month hindcasting survival analysis (Table~\ref{tab:predictive}). While Activity alone provides a strong baseline for survival prediction ($\text{AUC} = 0.81$)—reflecting the fact that inert systems are easily classified—it yields false positives for ``zombie'' ecosystems exhibiting high churn but structural decay. The inclusion of structural persistence ($\Gamma$) significantly enhances predictive accuracy ($\text{AUC} = 0.88$; $\Delta\text{AUC} = +0.07$, Wilcoxon $p < 0.05$). This indicates that while activity maintains the system's present state, it is the coupling with structural persistence that secures its future, effectively distinguishing sustainable autonomy from chaotic fragility.

\begin{table}[htbp]
\centering
\caption{Predictive Power Comparison}
\label{tab:predictive}
\begin{tabular}{lllll}
\toprule
\textbf{Model Feature} & \textbf{Metric Definition} & \textbf{AUC (12-month)} & \textbf{Improvement} & \textbf{Interpretation} \\
\midrule
Control & Project Age ($t$) & 0.47 & — & Non-predictive. Age alone does not \\
& & & & guarantee future survival (random \\
& & & & guess $\approx 0.5$). \\
\addlinespace
Baseline & Activity Only ($A_{\text{norm}}$) & 0.81 & Reference & Strong Baseline. High activity \\
& & & & correlates with survival, but yields \\
& & & & false positives for chaotic ``zombie'' \\
& & & & projects. \\
\addlinespace
Viability & $V = A_{\text{norm}} \times \Gamma$ & 0.88 & +0.07 & Discriminative. The interaction of \\
& & & & structure and activity significantly \\
& & & & outperforms activity alone \\
& & & & (Wilcoxon $p < 0.05$). \\
\bottomrule
\end{tabular}
\end{table}

External validation confirmed that V discriminates governance tiers (Kruskal-Wallis $H=9.51$, $p<0.01$), while $\Gamma$ alone does not ($p=0.68$) due to structural fossilization in abandoned projects. Governance tier explains approximately 20\% of variance in viability ($\eta^2 = 0.20$), with both Corporate and Foundation-governed ecosystems significantly outperforming Personal projects ($p < 0.003$), confirming that V captures active sociotechnical life rather than inert persistence.

\subsection*{Methodological Robustness}

The construct validity of $\Gamma$ was verified against alternative aggregation functions (Harmonic Mean, Geometric Mean), yielding Spearman correlations $r > 0.96$. Furthermore, the emergence of causal symmetry shows sensitivity to phase boundary selection. While robust at 20\%, 33\%, and 50\% cut-points, the signal weakens or inverts at intermediate thresholds (25\%, 40\%), suggesting the transition follows a punctuated rather than linear progression.

\subsection*{Unexpected Findings}

Three results deviate from classical models of software evolution \cite{lehman1996} and merit emphasis.

First, the transition zone is crossed remarkably fast. Contrary to gradual maturation models, systems traverse the unstable intermediate regime in a median of 1.0 month—suggesting a critical transition rather than continuous drift. Once initiated, the phase shift completes rapidly.

Second, mature systems are not frozen. Fully 41\% of projects experience temporary regressions after reaching high-$\Gamma$ states. This refutes a ``crystallization'' model where mature architectures become rigid. Instead, autonomy requires continuous energetic maintenance against entropic drift—a dynamic homeostasis rather than static equilibrium.

Third, we observe what might be termed an Inverse Conway Effect. Conway's Law predicts that software structure mirrors team organization \cite{conway1968}. Our Granger analysis suggests the reverse also holds at maturity: architectural constraints increasingly shape coordination patterns, which then reinforce architectural boundaries. This bidirectional lock-in was not predicted by classical software evolution theory.

\subsection*{Summary of Hypothesis Tests}

\begin{table}[htbp]
\centering
\caption{Summary of Hypothesis Tests}
\label{tab:hypothesis_summary}
\begin{tabular}{llll}
\toprule
\textbf{Hypothesis} & \textbf{Criterion} & \textbf{Result} & \textbf{Verdict} \\
\midrule
H1 (Primary) & Coupling ratio $\to 1$ & $0.65 \to 0.94$ & Supported \\
H2 & Two-regime GMM ($\Delta\text{BIC} > 0$) & $\Delta\text{BIC} = 1{,}850$ & Supported \\
H3 & Variance ratio $> 1.5$ & $1.70\times$ & Supported \\
H4 & \% reaching $\Gamma > 0.7$ exceeds 90\% & 100\% & Supported \\
H5 & Traverse time $< 6$ months & 1.0 month & Supported \\
H6 & Regression rate $< 50\%$ & 41\% & Supported \\
\bottomrule
\end{tabular}
\end{table}

\section*{Discussion}

Our results reveal that mature software ecosystems exhibit a statistical signature—phase transition, variance collapse, and causal symmetrization—consistent with operational closure. However, the interpretation of these findings requires careful epistemological framing.

\subsection*{Epistemological Framing: Two Notions of Causality}

Granger causality measures predictive information transfer: X ``Granger-causes'' Y if past values of X improve prediction of Y beyond Y's own history. This is a statistical property of time series, not a claim about generative mechanisms.

Rosen's closure to efficient causation is categorically different. A system achieves closure when its organization produces the constraints necessary for its own persistence—when efficient causes are themselves outputs of the system's processes ($f \in \text{Range}(F)$). This is a structural property of the causal architecture, not a temporal correlation.

As Pearl's framework clarifies \cite{pearl2009}, Granger relations concern observational associations, while Rosennean closure concerns what generates what. One cannot derive structural causation from predictive association without additional assumptions. We therefore resist the direct inference from Granger symmetry to closure and summarize our interpretive stance in Table~\ref{tab:causality}.

\begin{table}[htbp]
\centering
\caption{Distinguishing Statistical Signatures from Systemic Claims}
\label{tab:causality}
\begin{tabular}{p{3.5cm}p{3.5cm}p{4cm}p{3.5cm}}
\toprule
\textbf{Empirical Observation} & \textbf{Dynamic Interpretation} & \textbf{Rejected Hypothesis} & \textbf{Inference Limit} \\
\midrule
Bimodality of $\Gamma$ (Hartigan dip, $\Delta$BIC) & Phase transition: shift from exploratory to sedimented regime & Linear continuum: software evolution is not simple progressive accumulation & Does not derive from a physical Master Equation \\
\addlinespace
Granger Symmetry ($A \to S$ becomes $A \leftrightarrow S$) & Bidirectional coupling: architecture constrains activity as much as it stems from it & External steering: system is not a pure passive artifact & Does not prove the system generates its own components (Rosen's closure) \\
\addlinespace
Variance Collapse ($\sigma^2$ reduced $1.70\times$) & Dynamic attractor: system maintains internal state despite perturbations & Random walk: structure does not drift randomly & Refers to information (Shannon) entropy, not physical/thermal laws \\
\bottomrule
\end{tabular}
\end{table}

\begin{figure}[H]
    \centering
    \includegraphics[width=0.75\linewidth]{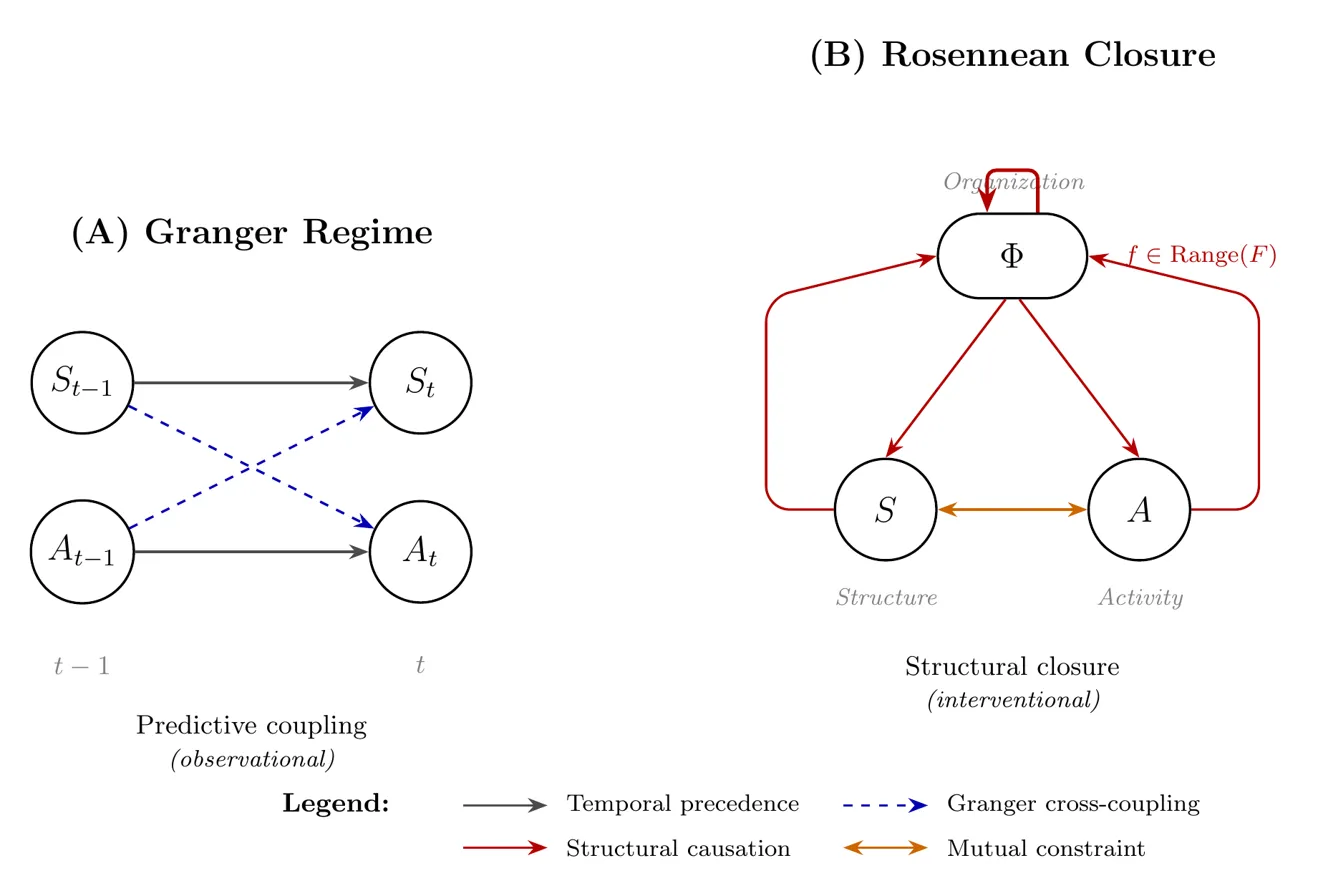}
    \caption{ Two notions of causality. (A) Granger regime: bidirectional predictability between Structure ($S$) and Activity ($A$) across time (observational). (B) Rosennean closure: the system's organization ($\Phi$) generates its own constraints ($f \in \text{Range}(F)$). Our tests address (A); inference to (B) requires additional mechanistic evidence.}
    \label{fig:casual1}
\end{figure}

\begin{figure}[H]
    \centering
    \includegraphics[width=0.75\linewidth]{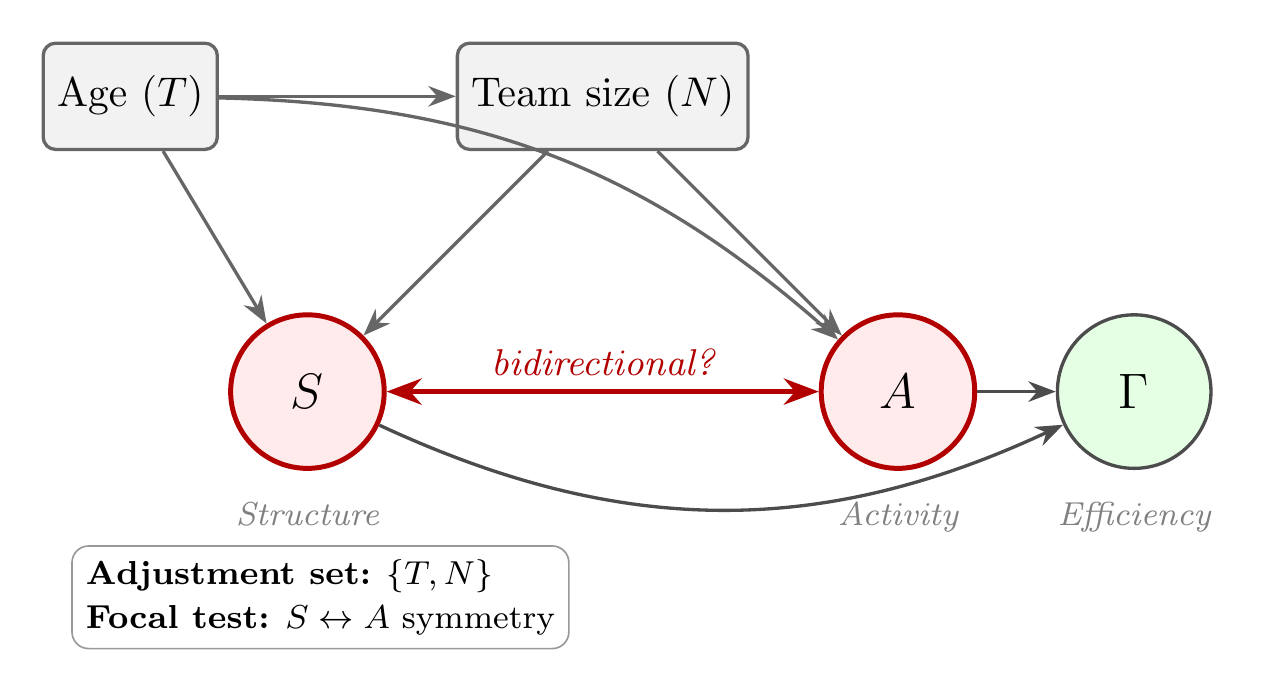}
    \caption{Causal Coupling Dynamics. Empirical model derived from segmented Granger causality analysis. In the emergent phase, Activity predominantly drives Architecture ($A \to S$). At maturity, Architecture reciprocally constrains Activity via established interfaces ($A \leftrightarrow S$). The convergence of the coupling ratio from 0.71 to 0.94 constitutes the operational signature of closure—a necessary but not sufficient condition for Rosennean closure.}
    \label{fig:casual}
\end{figure}

The finding that 41\% of mature projects experience regressions refutes a naive ``frozen state'' hypothesis. It suggests instead a dynamic homeostasis model, where the system must actively expend energy to maintain the high-$\Gamma$ state against entropic drift. This observation motivates the question: what is the epistemic status of the symmetrization we observe?

\subsection*{Granger Symmetry as a Necessary Condition}

We propose that Granger symmetrization constitutes a necessary condition for operational closure, though not a sufficient one. The argument proceeds by contraposition: if a system exhibits genuine closure, bidirectional statistical coupling should obtain. Extreme asymmetry—where structure entirely determines activity (external imposition) or where activity entirely determines structure (passive material)—is incompatible with closure.

Our data show that mature ecosystems converge toward symmetric coupling, eliminating the hypothesis that these systems remain purely activity-driven artifacts. This is a diagnostic rather than demonstrative contribution: we can rule out a class of explanations (external causation, passive persistence) rather than definitively establish closure.

\subsection*{The Sociotechnical Mechanism}

If Granger symmetry is a necessary signature, what generates it? Covariate control analysis indicates that bidirectional coupling retains approximately 70\% of its effect size when normalizing by contributor count, suggesting the signal is predominantly architectural rather than a team-size artifact. We interpret this through what might be termed the Inverse Conway Effect. Conway's Law predicts that software mirrors team structure; our data suggest that at maturity, the relationship becomes reciprocal: architecture constrains coordination patterns, which constrain future architecture.

While Granger analysis establishes statistical coupling, the concrete mechanisms through which architecture constrains activity are observable in modern development workflows. Three pathways implement this constraint:

\begin{enumerate}
     \item \textit{Structural coupling}: The dependency graph itself constitutes the primary constraint. Class inheritance hierarchies, method signatures, and import relationships create a causal topology where modifications to central nodes propagate effects throughout the system. Unlike process-level enforcement, these constraints are ontological rather than social—a developer can disable a linter but cannot ignore a compilation error. As the codebase matures, the space of viable modifications narrows: new code must conform to existing type contracts, interface definitions, and coupling patterns simply to function.
 
    \item \textit{Automated gating}: CI/CD pipelines and test suites encode the system's historical invariants, automatically rejecting contributions that violate established patterns—often without human intervention.
    
    \item \textit{Static enforcement}: Linters, style checkers, and static analyzers compel developers to conform their output to the codebase's structural template, effectively ``training'' new contributors to maintain organizational coherence.
    
    \item \textit{API dependency}: As the stable core (high-$\Gamma$ code) grows, new features must integrate through existing interfaces. The cost of violating these architectural boundaries becomes prohibitive, channeling developer activity through pre-existing structural pathways.
\end{enumerate}

These mechanisms constitute the selective membrane through which the system filters developmental inputs. While we do not claim this proves Rosennean closure, it demonstrates that the statistical signature we observe corresponds to concrete, identifiable enforcement mechanisms in software practice. This bidirectional constraint—if confirmed mechanistically—would satisfy Rosen's criterion at the coupled sociotechnical level; however, alternative explanations including shared external dependencies cannot be excluded without further mechanistic studies.

\subsection*{Resolution of the Mirroring Anomaly}

Our findings provide a mechanism for the Conway's Law violation documented by Colfer and Baldwin \cite{colfer2016}. In hierarchical firms, managerial coordination imposes architectural boundaries---structure mirrors organization because organization \textit{enforces} structure. Open-source systems lack this imposition yet maintain coherent architecture.

We propose that causal symmetrization provides the functional substitute. As ecosystems mature, the architecture itself assumes the constraining role that managers play in firms. The bidirectional coupling we observe (ratio $\to 0.94$) means that established interfaces increasingly filter contributions, enforcing boundaries without hierarchical authority. Open-source systems do not violate Conway's Law---they satisfy it through a different mechanism: architectural constraint replaces organizational constraint (Fig. 6).

\subsection*{Scope of Claims}

We distinguish three levels of assertion:

\begin{enumerate}
    \item \textit{Empirical} (strong confidence): Mature ecosystems exhibit phase transition, variance collapse, and Granger symmetrization. This is reproducible across 50 systems.
    \item \textit{Interpretive} (moderate confidence): Symmetrization constitutes an operational signature consistent with—and necessary for—closure. Systems lacking this signature can be excluded as candidates for operational autonomy.
    \item \textit{Theoretical} (provisional): Mature sociotechnical systems may instantiate closure dynamics functionally analogous to those in living systems, suggesting closure is substrate-independent.
\end{enumerate}

The contribution of this work is not to prove that software ``lives,'' but to demonstrate that the question is empirically tractable—that theoretical concepts from organismic biology can be operationalized in artificial systems, yielding non-trivial discriminations.

\subsection*{Connections to Broader Research Programs}

Our findings connect to several active research fronts beyond software engineering. 
The variance collapse we observe ($1.77\times$) prior to dynamic stability inverts the classical `critical slowing down' signature---increasing variance before collapse---identified by Scheffer et al.\ in ecological systems \cite{scheffer2009}. This symmetry suggests that variance dynamics may serve as bidirectional indicators: expansion signals approaching instability; contraction signals approaching autonomy. By demonstrating these signatures in software ecosystems, we respond to Scheffer et al.'s call to test critical transition frameworks beyond ecology \cite{scheffer2012}, suggesting substrate-independent principles of complex system dynamics.

The bidirectional coupling between structure and activity aligns with network controllability theory, which distinguishes driver nodes from redundant ones in directed networks \cite{liu2011}. Our results suggest that mature sociotechnical systems develop distributed controllability—no single node (or developer) steers the system; control emerges from architectural constraints.

Finally, our framework extends computational social science methods \cite{lazer2009} toward questions of collective autonomy. While most large-scale studies of collaboration focus on productivity or network structure, we introduce dynamics of self-organization as a measurable outcome—potentially applicable to scientific collaborations, open governance systems, and decentralized organizations.

\section*{Limitations}

Several constraints bound our conclusions. First, the Gamma metric is a heuristic proxy for metabolic efficiency, not a direct physical measurement. While sensitivity analysis ($\lambda \in [0.6, 1.0]$) confirms robust topological signal, the specific scalar values depend on the chosen penalty function. Similarly, the high correlations ($r > 0.99$) between Gamma formulations reflect the mathematical proximity of multiplicative aggregation functions rather than unexpected empirical precision.

Second, the causal symmetrization signal shows moderate sensitivity to phase boundary selection (robust for 3/5 tested cut-points), warranting adaptive segmentation methods in future work. Third, while $N = 50$ ecosystems exceeds typical MSR studies, generalization to proprietary systems remains untested. Fourth, Granger causality establishes predictive precedence, not generative mechanism.

Finally, extremely high Z-scores ($> 25$) partly reflect large sample size ($N \approx 10^4$ months); we therefore emphasize effect sizes (Cohen's $d = 3.01$, $\eta^2 = 0.20$) over p-values for substantive interpretation.

We use biological terms (metabolism, homeostasis) as functional isomorphisms, not ontological claims. Software lacks the physicochemical substrate of biological life. However, our data suggest that in the informational domain, sociotechnical systems satisfy the formal criteria for closure to efficient causation defined by Rosen, independent of their material substrate.

\subsection*{Future Directions}

Moving from operational signature to mechanistic validation would require several complementary approaches: (1) Architectural decision tracing—qualitative analysis tracing specific constraints to their enforcement mechanisms through code review processes and architectural decision records; (2) Perturbation studies—case studies of major refactoring events (Linux's transition from BitKeeper to Git, Kubernetes' API deprecation cycles) to reveal whether structural recovery follows intrinsic constraints or external intervention; and (3) Counterfactual modeling—agent-based simulations instantiating bidirectional constraint reproduction to generate predictions testable against empirical trajectories. We view the present study as establishing statistical preconditions for closure; mechanistic verification remains an open program.

\section*{Conclusion}

We have identified causal symmetrization as an empirically validated signature of operational autonomy in complex collaborative systems. Analyzing 50 ecosystems over 10,000 system-months, we find that mature systems converge toward a symmetric coupling regime (ratio $0.65 \to 0.94$) where organizational structure and collective activity mutually constrain each other. This phase transition, accompanied by variance collapse and attractor dynamics, satisfies all six pre-registered hypotheses.

These results establish a general principle: large-scale collaborative systems can develop self-organizing dynamics functionally analogous to those theorized in biological contexts—not through biological mechanisms, but through the emergence of bidirectional constraint between structure and activity. The signature we identify is substrate-independent and potentially applicable across complex adaptive systems where organizational architecture and collective behavior co-evolve.

We propose causal symmetrization as a quantitative criterion for operational autonomy, applicable beyond software to scientific collaborations, open governance systems, decentralized organizations, and hybrid human-AI collectives. The methodological framework developed here—combining structural persistence metrics with directional causality analysis—offers a transferable approach for studying self-organization wherever complete observational data is available.

\section*{Data and Code Availability}

The complete dataset, including the topological analysis engine and the replication scripts used in this study, are available in the Zenodo repository: \cite{gosme2025}.

\section*{Author contributions}
A.G. designed the study, performed the analysis, and wrote the manuscript.

\section*{Competing interests}
The author declares no competing interests.
\section*{Acknowledgements}
The author thanks the open-source communities whose collaborative 
work made this analysis possible.


\bibliographystyle{naturemag} 
\bibliography{references}     

\end{document}